\documentclass[twocolumn]{jpsj2} 
\title{Spin-Resolved Edge States around an Antidot in the Vicinity of the $\nu =2$ Quantum Hall State}

\author{Masanori \textsc{Kato}%
, Akira \textsc{Endo}\thanks{E-mail address: akrendo@issp.u-tokyo.ac.jp}, Shingo \textsc{Katsumoto}, and Yasuhiro \textsc{Iye}}

\inst{Institute for Solid State Physics, University of Tokyo, 5-1-5 Kashiwanoha, Kashiwa, Chiba 277-8581}

\abst{We have investigated spin-related effects on the electronic transport in a single antidot system in the vicinity of the $\nu = 2$ quantum Hall state where two edge states with different spins of the lowest Landau level are formed around the antidot. The conductance exhibits the paired $h/2e$ 
Aharonov-Bohm (AB) oscillations. 
Using a tilted magnetic field technique, we have observed the evolution of the oscillations with the total magnetic field and a concomitant change in the source-drain bias dependence of the differential conductance. From the observation, we extract the effect of the Zeeman energy on the AB oscillations.}

\kword{GaAs/AlGaAs, two-dimensional electron system, quantum Hall effect, antidot, Aharonov-Bohm effect, Zeeman effect}

\begin{document}
\maketitle

\section{Introduction}
Transport properties of the quantum Hall (QH) states~\cite {KlitzingIQH} are succinctly interpreted in terms of the so-called edge channel picture~\cite {Prange, Yoshioka}. 
At integer Landau level fillings, current is carried by dissipationless edge channels formed at the sample boundaries. 
The edge channels have chirality, i.e., they can pass the current only in one direction. 
Dissipative transport occurs either when the filling factor $\nu$ is away from an integer and thus the current is carried by extended bulk states of the partially occupied topmost Landau level (LL), or when electrons can be transferred to the counterpropagating edge channels on the opposite side of the sample as in the case of a quantum wire narrow enough to allow such transfer by quantum tunneling. 
The edge channel picture gives an intuitively appealing description of the QH state.
For non-interacting electrons, the edge channels are envisaged simply as occurring at the crossing of the Landau levels with the Fermi level.
However the electron-electron interaction fundamentally changes the picture.
Chklovskii \textit{et al.}~\cite {Chklovskii} have shown that alternating bands of compressible and incompressible state are formed in the edge region of a QH system.
The effective potential is concomitantly modified from the bare confining potential in a significant way.

A quantum antidot, which is an artificial potential hill tailored in a two-dimensional electron system (2DES), provides a fascinating stage to study the QH edge states~\cite {Sachrajda, Goldman1995}. 
Under a strong perpendicular magnetic field, electrons form bound states around an antidot, giving rise to a series of single-particle (SP) states with discrete energies.
Each SP state encloses an integer multiple of magnetic flux quanta.
As the perpendicular magnetic field is adiabatically changed, the SP states readjust themselves to keep the enclosed flux constant~\cite {SimReview}.
Namely, the radius of the circular SP state decreases (increases) with increasing (decreasing) perpendicular field, and accordingly the SP energy increases (decreases) reflecting the negative slope of the antidot potential.
Successive crossing of the SP states with the Fermi level ($E_\mathrm F$) gives rise to Aharonov-Bohm (AB) oscillations in various physical quantities.

The edge states around an antidot (termed hereafter \lq\lq antidot states") can be accessed, for example, by placing the antidot in the middle of a constriction so as to make them couple to the extended edge channels.
Tunneling via antidot states then provides a transport channel between the edge channels on the both sides of the sample, and the magnetoconductance of the system exhibits the AB oscillations. 
Although the basic features of the oscillating magnetoconductance such as the oscillation period can be understood within the simple non-interacting electron picture, many experimental~\cite{Fordsingle, ChargeKataoka, Kataokadf, Kataokakondo, Maasilta, GoldmanIQH} and theoretical \cite{SimKondo, Hwanganti, ZozouSD} studies have revealed the crucial role of the electron-electron interaction. 

Let us focus on the case of $\nu_\mathrm{c}=2$, $\nu_\mathrm{c}$ being the local Landau level filling in the region of the constriction (around the antidot), where two edge states with different spins of the lowest Landau level are formed around the antidot.
Previous studies in this regime have revealed AB oscillations that have a paired structure~\cite{Mace, Michael, Kato, Bassett}.
This characteristic AB oscillation waveform is basically interpreted in terms of the energy spectrum of the antidot states which are governed by the following two energies; the energy spacing between two successive SP states with the same spin $\Delta E_\mathrm{sp}$ and the Zeeman splitting $E_\mathrm{Z}$. 
The magnetic-field dependences of them are given as 
\begin{align}
\Delta E_\mathrm {SP} &=\frac {h}{2\pi er B_\perp } \left |\frac {\mathrm d E_\mathrm {AD} (r)}{\mathrm d r} \right |, \label {eq:dEad}\\
E_\mathrm Z &=g\mu _\mathrm B B,
\end{align}
where $r$, $B_\perp $, and $|\mathrm d E_\mathrm {AD} (r)/\mathrm d r|$ are the average radius of the two SP states with different spins, the perpendicular component of the total magnetic field $B$, and the slope of the potential hill forming the antidot, respectively. 
In the noninteracting model, $|\mathrm d E_\mathrm {AD}/\mathrm d r|$ is independent of magnetic field, so $\Delta E_\mathrm {SP}$ is inversely proportional to the perpendicular field component, that is $\Delta E_\mathrm {SP} \propto 1/B_\perp $.  
On the other hand, the Zeeman energy is proportional to the total magnetic field $E_\mathrm Z \propto B$. 

Previous studies have also uncovered interesting phenomena such as double frequency ($h/2e$) AB oscillations~\cite {Fordsingle, Kataokadf} and a Kondo-like behavior~\cite {Kataokakondo}, that obviously call for interpretations based on a more realistic model for the edge states around the antidot than the non-interacting picture.
Indeed, it has been suggested that a fully compressible edge states may not be formed around an antidot owing to the strong interaction in a finite quantum system~\cite {Karakurt}. 

In this work, we investigate the AB oscillations of an antidot system around the $\nu_c =2$ QH state, their behavior as a function of magnetic field and source-drain bias.
Tilted field technique enabled us to separate the effect of the Zeeman splitting from that of the orbital effect.
We find a marked difference in the behavior of the antidot states between the lower and higher field sides of the $\nu_\mathrm{c} =2$ QH state.

\section{Experimental}

\begin{figure}[tb]
\begin{center}
\includegraphics[width=0.9\linewidth]{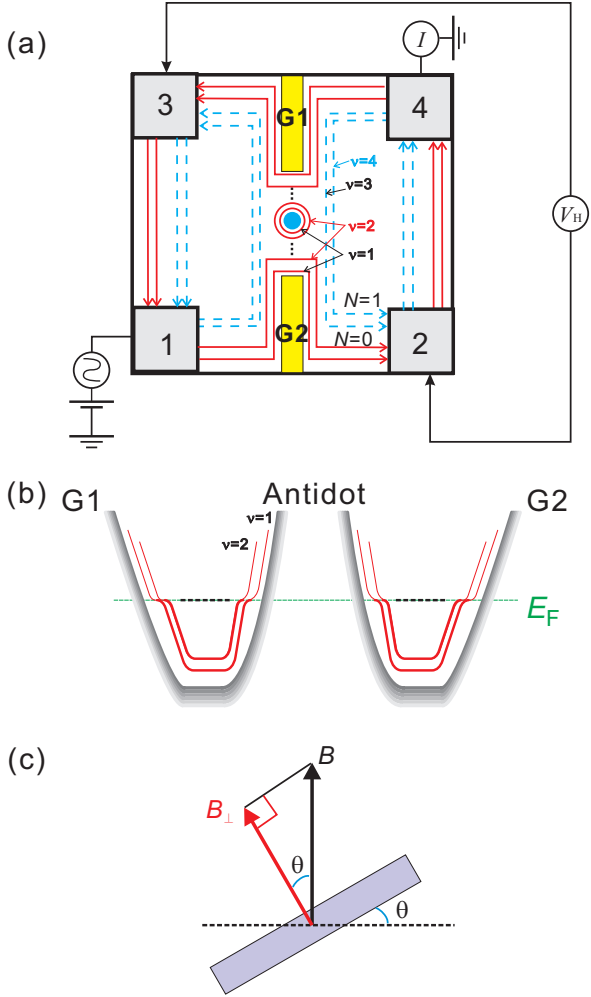}
\caption{(Color online) (a) Illustration of  the Van der Paw sample. Numbered rectangles are Ohmic contacts. The arrowed lines show edge channels. (b) Schematic energy diagram across the constrictions (the cross section along the line connecting G1 and G2). Solid red lines represent $N=0$ Landau levels with thicker lines denoting the occupied states.
(c) Geometry of the tilted field technique. The sample is tilted by an angle $\theta $ in a vertical magnetic field $B$.}
\label{FIG1}
\end{center}
\end{figure}

The antidot sample used in the present study was fabricated from a GaAs/AlGaAs single heterojunction wafer with 2DEG density $3.8 \times 10^{15}~\mathrm{m^{-2}}$ and mobility $60~\mathrm {m^2/Vs}$.  
Standard techniques of electron beam lithography, wet chemical etching and metal evaporation were used to fabricate the device.
A circular antidot of radius 180~nm was placed in the middle of the gap between two 400~nm-wide Schottky gates (G1, G2), as illustrated in Fig.~\ref{FIG1}(a).
The G1 and G2 gates were biased independently so as to adjust the effective distance between the extended edge channel(s) and the localized edge channel(s) around the antidot.
Under a large negative bias, the filling factor in the constriction region $\nu_\mathrm{c}$ becomes less than that in the bulk of the sample $\nu_\mathrm{b}$. 

The conductance $G$ across the constriction was measured by the following method so as to eliminate the effect of series resistance. 
A $5~\mathrm {\mu V} $ AC (77 Hz) excitation voltage was applied to ohmic contact 1 as shown in Fig.~\ref {FIG1}(a). 
The contact 4 was connected to a virtual ground through a current amplifier and the current $I$ passing through this circuit was measured. 
With a simultaneously measured Hall voltage $V_\mathrm H$ between contacts 2 and 3, the sample conductance was defined as
\begin{equation}
G=\frac {I}{V_\mathrm H} = \nu _\mathrm c \frac {e^2}{h}. \label {eq:Gad}
\end{equation}

In the QH regime, this is effectively equal to an ideal two-terminal conductance of the constriction (i.e., with the series resistance eliminated). 
This feature comes from the chiral and adiabatic nature of the edge channel transport, which does not produce any voltage drop between the probes 3 and 1, or between 2 and 4. 
A DC source-drain voltage $V_\mathrm {SD}$ was applied to investigate the conduction under a finite bias.
The DC bias was also used to cancel an offset voltage of the order $\sim 100~\mathrm {\mu V}$ that appeared across the device. 

The sample was cooled to 50 mK in the mixing chamber of a top-loading dilution refrigerator, and a magnetic field was applied by a 15~T superconducting solenoid. 
The top-loading probe was equipped with a rotating stage so that the magnetic field angle $\theta$ from the plane normal could be changed \textit{in situ}, as illustrated in Fig.~\ref {FIG1}(c).

\section{Results and Discussion}
\subsection{AB oscillations in the vicinity of $\nu_\mathrm c =2$} \label {secobsingle}

\begin{figure}[tbp]
\begin{center}
\includegraphics[width=0.9\linewidth]{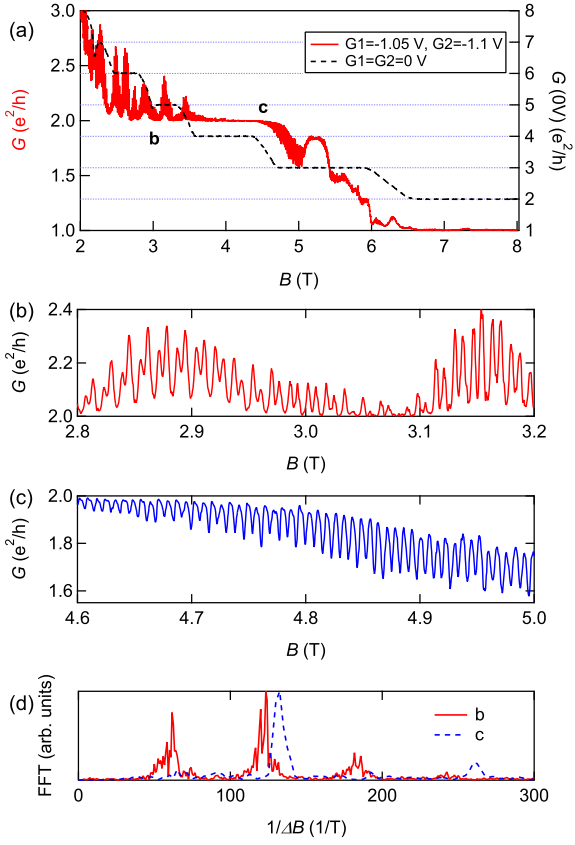}
\caption{(Color online) (a) Solid red line: magnetic-field dependence of the antidot conductance $G$ around $\nu _\mathrm c =2$ (left axis). Dashed black line: magnetoconductance of the bulk 2DES without applying gate voltages (right axis). The source-drain bias $V_\mathrm {SD}=70~\mathrm {\mu V}$ is applied to cancel the offset voltage. 
(b),(c) Expanded $G$ traces for the field range on the lower and higher field sides of the $\nu _\mathrm c=2$, respectively, as labeled in (a). 
(d) The Fourier power spectra of the oscillatory components at \textbf b and \textbf c. }
\label{FIG2}
\end{center}
\end{figure}

Figure~\ref{FIG2}(a) shows the conductance $G$ through the antidot in the vicinity of $\nu_\mathrm{c} = 2$, where two spin-split edge states of the lowest LL encircle the antidot while higher LL edge channels are reflected by the constriction.
(The bulk filling $\nu_\mathrm{b}$ ranges from 2 to 8 over this field range, as seen from the conductance under zero gate bias shown by the dashed curve).  
Figures \ref{FIG2}(b) and (c) are expanded traces on the lower and higher field side of $\nu_\mathrm{c} = 2$, respectively.
On the low-$B$ side (Fig.~\ref{FIG2}(b)), a train of paired peaks with alternating heights appear above the $2e^2/h$ plateau. 
On the high-$B$ side (Fig.~\ref {FIG2}(c)), by contrast, successive dips that appear below the $2e^2/h$ plateau are about the same depth. 
The Fourier power spectra (Fig.~\ref {FIG2}(d)) make the difference more apparent; i.e., split peaks for the lower field data (solid red curve) vs. a single $h/2e$ peak for the higher field data (dashed blue curve). 
It is noted that the Fourier peak position for the high-$B$ side data is close to the position of the split second harmonic peak for the low-$B$ side data, and that the former is slightly shifted toward the right (higher frequency side) as compared to the latter.
Namely, the AB oscillation frequency on the high-$B$ side is twice the fundamental frequency $1/\Delta B=e\pi r^{*2}/h$ corresponding to the antidot area.
The slightly higher frequency reflects the fact that the relevant area of the outer edge (i.e., the spin-down lowest LL $(0, \downarrow)$ state) is larger on the high-$B$ side, where this edge state is on the verge of delocalization. 

As the magnetic field is increased further ($B \geq 5.5~\mathrm T$), the conductance $G$ approaches the $e^2/h$ plateau, and the AB oscillations turn to those with the fundamental frequency $1/\Delta B=e\pi r^{*2}/h$.
The AB oscillations in this range are attributed to resonant transmission by spin-conserving inter-LL scattering, the (3-1) tunneling, through the lowest LL from the second LL\@. Here we used the notation $(\nu_\mathrm{b}$-$\nu_\mathrm{c})$ to denote the tunneling from the bulk $\nu_b$ extended edge state to $\nu_\mathrm{c}$ edge state localized around the antidot. 
The AB period $\Delta B=16~\mathrm {mT}$ corresponds to an effective antidot radius $r^* \sim 280 \mathrm{nm}$, which is in reasonable agreement with the lithographical radius $r=180~\mathrm {nm}$ plus the width ($\sim 100~\mathrm{nm}$) of the depletion region.
Similar features were observed around the $\nu_\mathrm c =2$ plateau for different settings of the gate voltage and magnetic field. 

In the following subsections, the features of the AB oscillations on the low-$B$ and high-$B$ sides of the $\nu_\mathrm c=2$ plateau will be discussed in more detail, with particular attention to the effect of Zeeman splitting. 

\subsection{Low-$B$ side of $\nu_\mathrm c =2$}

\begin{figure}[tbp]
\begin{center}
\includegraphics[width=0.95\linewidth]{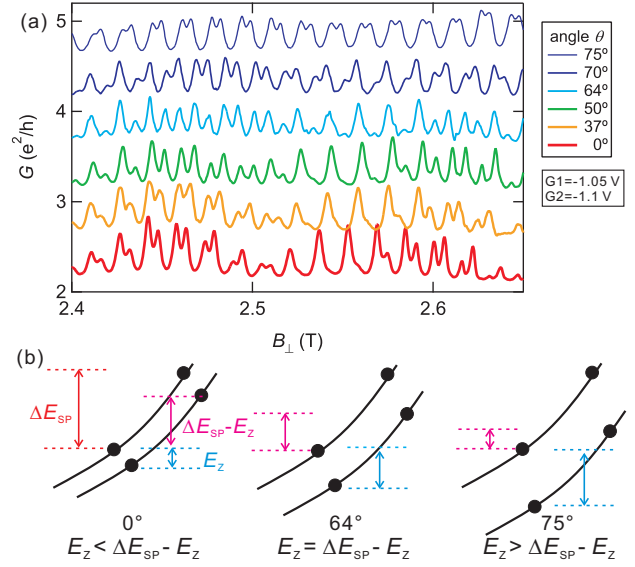}
\caption{(Color online) (a) The sample conductance $G$ across the antidot as a function of perpendicular field component $B_\perp$ on the lower field side of $\nu_\mathrm {c}=2$ ($\nu_\mathrm{b}=6$) 
for different tilt angles $\theta $. The successive traces are offset by $0.5~e^2/h$. 
The source-drain bias $V_\mathrm {SD}=70~\mathrm {\mu V}$ is applied to cancel the offset voltage. 
(b) Schematic drawings of the spin-split lowest Landau levels around the antidot in the single-particle model. Solid circles represent SP energy levels.
While $\Delta E_\mathrm {SP}$ is unchanged for the same $B_\perp $, $E_\mathrm {Z}$ increases proportional to the total field $B$.}
\label{FIG3}
\end{center}
\end{figure}

Figure~\ref{FIG3}(a) shows the conductances $G$ through the constriction under the condition $\nu_\mathrm{b}=6$ and $\nu_\mathrm{c}=2$ as a function of the perpendicular field component $B_\perp $ for different tilt angles $\theta$. 
The lowermost trace for $\theta=0^\circ $ exhibits AB oscillations that occur as a series of double peaks rising above an approximately constant baseline at $\sim 2e^2/h$. 
The twin-peak structure is a manifestation of partial transmission of the second lowest LL edge channels via tunneling into (and out of) the lowest LL edge states localized around the antidot, namely, ($\nu_\mathrm{b}$-$\nu_\mathrm{c}$)$=$(3-1) and (4-2) tunnelings. 
This picture is corroborated by the fact that the twin-peak structure is not observed for low bulk fillings $\nu_\mathrm{b} \leq 3$.  

As the sample is tilted, the pattern of the AB oscillations changes.
The evolution is most clearly observed in the traces around $B_\perp\simeq 2.45~\mathrm T$:
the spacing between the twin peaks increases with increasing $\theta $, until the peaks become equally spaced at $\theta = 64^\circ $.
Further increase of $\theta $ restores the twin-peak structure. 

This behavior can be understood by considering the spin-resolved single-particle spectrum as illustrated in Fig.~\ref{FIG3}(b). 
For a given $B_\perp$, the Zeeman splitting $E_\mathrm{Z}$ increases with increasing $\theta $.
As shown in Fig.~\ref{FIG3}(b), the energy spacing between successive states alternates between $E_\mathrm{Z}$ and $\Delta E_\mathrm{SP} - E_\mathrm{Z}$.
At $\theta =0^\circ$, $E_\mathrm {Z}$ is smaller than $\Delta E_\mathrm {SP} - E_\mathrm {Z}$. 
As $\theta $ is increased, $E_\mathrm{Z}$ increases and coincides with $\Delta E_\mathrm{SP} - E_\mathrm{Z}$ at some point.
In Fig.~\ref{FIG3}(a), this occurs at $\theta \simeq 64^\circ$.
When $\theta $ is increased further, $E_\mathrm {Z}$ becomes larger than $\Delta E_\mathrm {SP} - E_\mathrm{Z}$ and the pair structure emerges again. 

\begin{figure}[tbp]
\begin{center}
\includegraphics[width=0.95\linewidth]{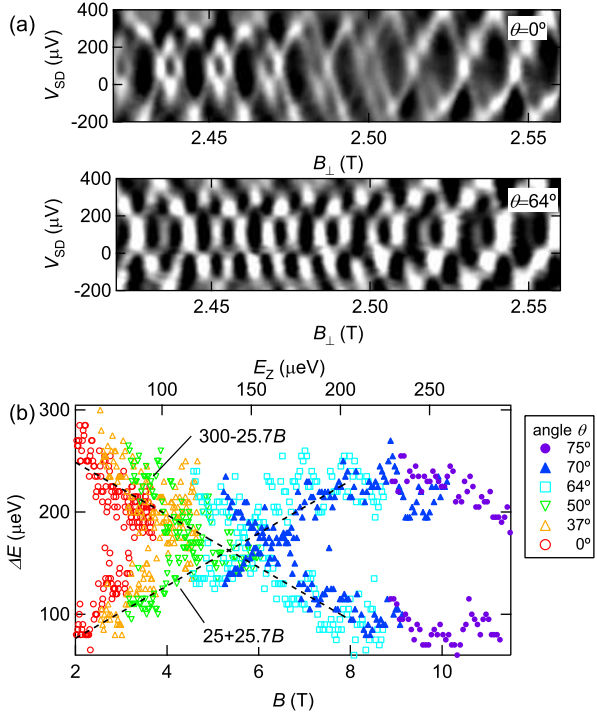}
\caption{(Color online) (a) Gray-scale plots of the oscillatory component of differential conductance as a function of $B_\perp$ and $V_\mathrm {SD}$. 
The lighter regions indicate the higher (peak) conductance. The two figures are for the tilt angle $\theta =0^\circ $ (top) and $64^\circ $ (bottom), respectively. 
(b) Evolution of the energy splitting $\Delta E$ with the total field $B$ (the corresponding Zeeman splitting $E_\mathrm Z$ is also indicated in top axis), obtained from the DC-bias measurement. 
Different symbols present the energy spacings at different tilt angles. 
The dashed straight lines represent $B$-linear fit.}
\label{FIG4}
\end{center}
\end{figure}

In order to verify the above picture, we investigated the dependence of the differential conductance on $B_{\perp}$ and $V_\mathrm{SD}$. 
Figure~\ref {FIG4}(a) presents gray-scale plots of the differential conductance on a $B_\perp$-$V_\mathrm{SD}$ plane for $\theta=0^\circ$ and $64^\circ$. 
Each trace in Fig.~\ref{FIG3}(a) corresponds to the cross-sectional profile of the gray-scale plot along the horizontal line at an offset bias $V_\mathrm {SD}=70~\mathrm {\mu V}$ passing through the center of the diamonds. 
The diamond patterns in Fig.~\ref{FIG4}(a) reflects spin-resolved resonances of the antidot states. 
At $\theta =0^\circ$, the spin-resolved resonances are manifested by the series of diamonds alternating in size.
The vertical height of the diamond corresponds to the energy difference between the two successive states plus the charging energy, i.e., $\Delta E_\mathrm{SP} - E_\mathrm{Z} +e^2/C$ or $E_\mathrm{Z} +e^2/C$~\cite {ChargeKataoka, Michael}. 
At $\theta =64^\circ $, the diamonds become nearly uniform in size, which signals coincidence of the two energy spacings.

Figure \ref{FIG4}(b) shows the energy spacings obtained from the vertical height of the diamonds as a function of the total magnetic field $B$ for different tilt angles $\theta$.
The experimental data follow two dashed straight lines corresponding to $\Delta E_\mathrm{SP} - E_\mathrm{Z} + e^2/C$ (down-slope line) and $E_\mathrm{Z} + e^2/C$ (up-slope line). 
These lines are drawn with a slope $E_\mathrm{Z}(\mu\mathrm{eV}) = 25.7 B(T)$ using the $g$-factor $|g| = 0.44$ for bulk GaAs. 
The deviation of the data from the dashed lines is attributed to the $B_\perp$-dependence of $\Delta E_\mathrm {SP}$ and $e^2/C$.

\begin{figure}[tbp]
\begin{center}
\includegraphics[width=0.9\linewidth]{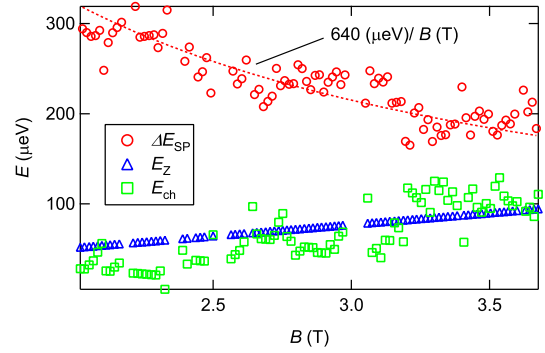}
\caption{(Color online) Dependence of energy gaps on the perpendicular magnetic field. Red circles, blue triangles, and green squares represent the energy spacing of the SP states 
$\Delta E_\mathrm {SP}$, the Zeeman splitting $E_\mathrm Z$, and the charging energy $E_\mathrm {ch}(=e^2/C)$, respectively. 
The data are obtained from $\Delta E$ of $\theta =0^\circ $ in Fig.~\ref {FIG4}(b). The dashed curve drawn through the $\Delta E_\mathrm {SP}$ is a fit to the $1/B$ dependence.}
\label{FIG5}
\end{center}
\end{figure}

Having determined the Zeeman splitting as $E_\mathrm{Z}(\mu\mathrm{eV}) = 25.7 B(T)$ using the standard $g$-factor $|g| = 0.44$, we can extract the values of $\Delta E_\mathrm{SP}$ and $e^2/C$ directly from the experimental values of $\Delta E_\mathrm{SP} - E_\mathrm{Z} +e^2/C$ and $E_\mathrm{Z} +e^2/C$ in Fig.~\ref{FIG4}(b). 
The $B_\perp$-dependence of $\Delta E_\mathrm{SP}$, $E_\mathrm{Z}$, and $E_\mathrm {ch}(=e^2/C)$ for $\theta=0^\circ$ are plotted in Fig.~\ref{FIG5}. 
The dashed curve drawn through the $\Delta E_\mathrm{SP}$ data represents a $1/B_\perp$-dependence. 

From the fit to Eq.~(\ref {eq:dEad}) we obtain the effective potential slope at the antidot edge $|\mathrm{d} E_\mathrm{AD}(r)/\mathrm{d} r|_{r^*} \simeq 2.7 \times 10^5~\mathrm {eV/m}$.
The good fit to the $1/B_\perp $-dependence indicates that the experimental results on the low-$B$ side of the $\nu _\mathrm c=2$ can be interpreted in terms of a single-particle picture with the screening effect. 

We comment on a similar experiment reported by Michael \textit{et al.}~\cite{Michael} for a smaller antidot sample.
(Their antidot radius 100~nm was smaller by a factor $\sim 2$ than the present one.)
They investigated the single-particle energy spacing $\Delta E_\mathrm {SP}$ and the charging energy, both deduced from the excitation spectrum of Coulomb diamonds, and found that $\Delta E_\mathrm {SP}$ decreased faster than the $1/B_\perp$-dependence. 
The seeming difference between the two experiments can be attributed to the relative importance of the Coulomb charging term. 
In the present case, $\Delta E_\mathrm{SP}$ is larger than the charging energy $E_\mathrm{ch}$. 
By contrast, in Michael \textit{et al.}'s experiment the deviation from the $1/B$-dependence of $\Delta E_\mathrm {SP}$ starts at the magnetic field where $E_\mathrm{ch}$  becomes larger than $\Delta E_\mathrm {SP}$.

\subsection{On the high-$B$ side of $\nu _\mathrm c =2$}

\begin{figure}[tb]
\begin{center}
\includegraphics[width=0.95\linewidth]{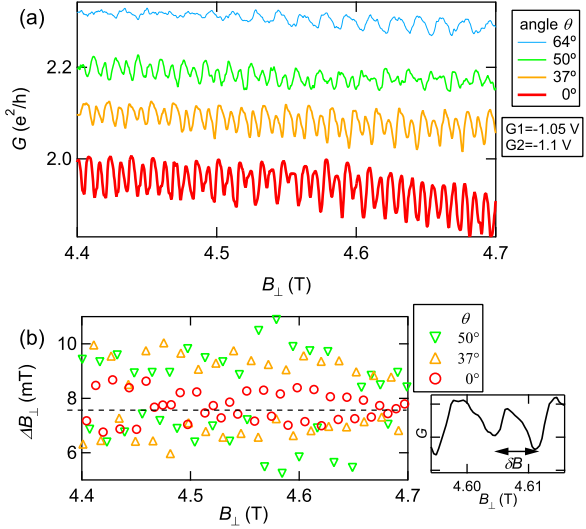}
\caption{(Color online) (a) The conductance $G$ across the antidot as a function of perpendicular field component $B_\perp$ on the higher field side of $\nu_\mathrm{c}=2$ 
for different tilt angles $\theta $. The successive traces are offset by $0.1~e^2/h$. 
The source-drain bias $V_\mathrm {SD}=70~\mathrm {\mu V}$ is applied to cancel the offset voltage. 
(b) The magnetic field spacing between adjacent dips at $\theta =0^\circ , 37^\circ , 50^\circ $. The dashed line indicates the oscillation period $\Delta B= 7.5~\mathrm {mT}$ 
obtained from the Fourier peak in Fig.~\ref {FIG2}(d). The inset shows a typical paired dip.}
\label{FIG6}
\end{center}
\end{figure}

Figure~\ref{FIG6}(a) shows the conductance $G$ through the constriction on the high-$B$ side of $\nu_\mathrm{c}=2$ as a function of $B_\perp $ for different tilting angles $\theta $. 
The bottom trace for $\theta =0^\circ$ exhibits dips with the average period $\Delta B\sim (h/2e)(1/\pi r^{*2})$ below the plateau $\sim 2e^2/h$.
To be precise, these dips are not equally spaced but are paired as shown in the inset of Fig.~\ref{FIG6}(b). 
These dips, which are observed in the filling range $2 \leq \nu_\mathrm{b} \leq  4$, represent intra-LL reflection resonances between the lowest LL edge channels.
At first thought, it seems natural to consider that these paired dips simply represent alternate backscattering resonances for the up-spin and down-spin channels.
However, as pointed out by Kataoka \textit{et al.}~\cite {Kataokadf, Kataokasel}, it is then difficult to understand why the two series of dips have comparative magnitude.
Since the tunneling distance should be significantly different between the up-spin and down-spin channels, the two series of dips should occur with vastly different magnitudes.
A more realistic picture presented by Kataoka \textit{et al.}~\cite {Kataokadf, Kataokasel} is that the intra-LL tunneling occurs only through the spin-down channel but the change 
in electron occupation of not only the spin-down but also the spin-up antidot states is reflected in the resonant reflection, giving rise to two resonances per the ordinary AB period.

\begin{figure}[tb]
\begin{center}
\includegraphics[width=0.95\linewidth]{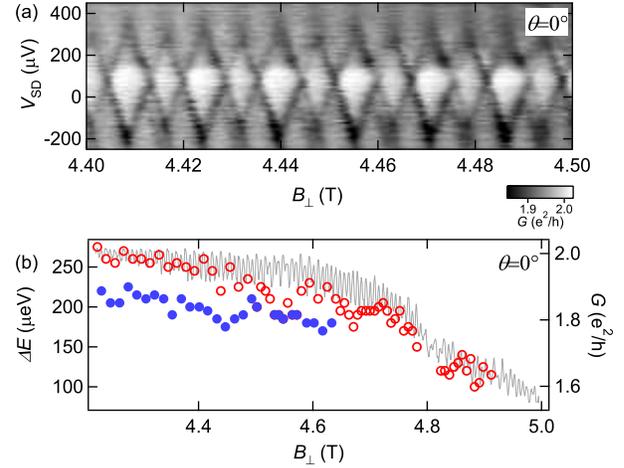}
\caption{(Color online) (a) Gray-scale plot of the differential conductance as a function of $B_\perp$ and $V_\mathrm{SD}$ at $\theta =0^\circ $. 
(b) The magnetic-field dependence of the energy gaps at $\theta =0^\circ $ (left axis). Open red and solid blue circles are the energy gaps obtained from the larger and smaller diamonds of (a). 
The gray line (right axis) indicates the magnetoconductance $G$ as a guide for eye. We suppose that open red and solid blue circles represent the energy scales 
$E_\mathrm {ch} + \Delta E^{(1)} _\mathrm {SP}$ and $E_\mathrm {ch} + \Delta E^{(2)} _\mathrm {SP}$, respectively.}
\label{dfE}
\end{center}
\end{figure}

As seen in Fig.~\ref{FIG6}(a), the paired dip structure evolves with the tilt angle $\theta$ until it becomes single period of $\Delta B \sim (h/e)(1/\pi r^{*2})$ at $\theta =64^\circ$. 
Figure~\ref{FIG6}(b) shows the magnetic field spacing between successive dips for different $\theta $, which oscillates between larger and smaller values, with the average value $(h/2e)(1/\pi r^{*2})$ (dashed line).
The difference between the two successive spacings increases with increasing $\theta$, and the smaller of the two ($\delta B$ shown in the inset of Fig.~\ref{FIG6}(b)) tends to vanish toward $\theta =64^\circ$. 

Figure~\ref{dfE}(a) is a gray-scale plot of the differential conductance on the $B_\perp$-$V_\mathrm{SD}$ plane for $\theta =0^\circ$ which manifests a train of diamonds with alternating size.
Figure~\ref{dfE}(b) shows the energy spacings $\Delta E$ obtained from the vertical height of the diamonds.
The red open circles are for the larger diamonds and the blue solid circles for the smaller ones.
At the lowermost of the present field range, the former is $\Delta E \sim 280~\mathrm {\mu eV}$ and the latter is $\sim 200~\mathrm {\mu eV}$ and they both decrease with increasing $B$.
For $B_\perp > 4.7~\mathrm{T}$, the pattern of alternating diamond size is smeared so that the data in this range are plotted with the same symbol.

\begin{figure}[tbp]
\begin{center}
\includegraphics[width=0.95\linewidth]{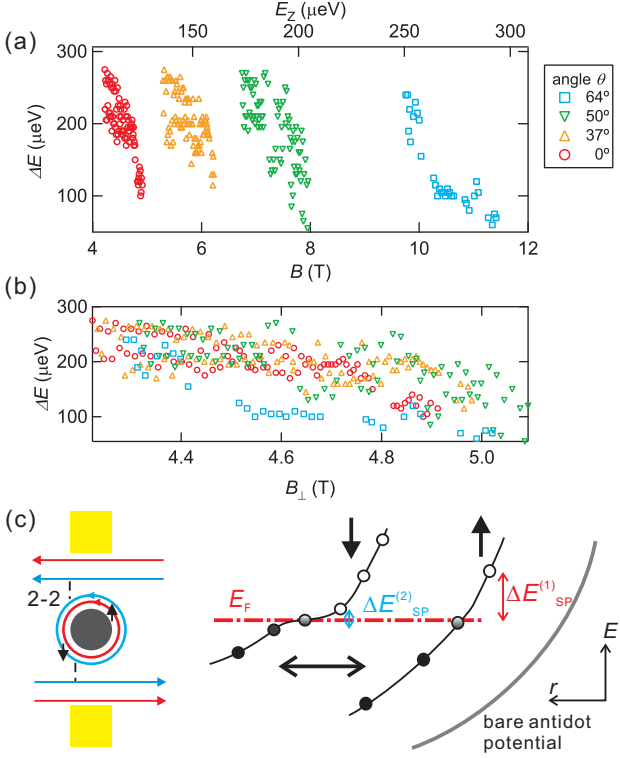}
\caption{(Color online) (a) Evolution of the energy splitting $\Delta E$ on the higher field side of $\nu _\mathrm{c}=2$ with the total field $B$ (bottom axis) and with the corresponding Zeeman splitting (top axis), obtained from the DC bias measurement. 
Different symbols represent the energy spacings at different tilt angles. 
(b) The plot of $\Delta E$ in (a) as a function of $B_\perp $. 
(c)  Left: Schematic picture of antidot and extended edge channels on the high-$B$ side of $\nu _\mathrm c=2$. The dashed lines indicate the (2-2) tunnelings. 
Right: Schematic picture of the spin-split lowest LLs around the antidot at the Fermi energy. 
Solid, open, and half-filled circles represent occupied, unoccupied, and partially occupied SP states, respectively. 
We propose that the $\nu =2$ (spin-down $\downarrow$) edge state is partially compressible and the $\nu =1$ (spin-up $\uparrow$) edge state 
remains incompressible. The separation between the two edge states changes with the Zeeman energy.}
\label{dfSD}
\end{center}
\end{figure}

Figure~\ref{dfSD}(a) shows the energy spacings (for both larger and smaller diamonds) obtained from the diamonds for different values of $\theta$.
These four sets of data look alike despite they are taken under substantially different total magnetic fields $B$.
Indeed, when plotted against $B_\perp$, these data nearly overlap with one another, as shown in Fig.~\ref {dfSD}(b).
It looks as if the Zeeman splitting $E_\mathrm{Z}$ does not contribute to the energy spacing $\Delta E$.
This result strongly contrasts with the behavior observed on the low-$B$ side.
It is evident that the single-particle picture which seems to work for the low-$B$ side is no longer valid on the high-$B$ side.

Let us discuss this anomalous behavior of the AB oscillations in more detail. 
We interpret our results based on the picture proposed by Ihnatsenka and Zozoulenko~\cite{ZozouSD}.
Their calculations based on density functional theory (DFT) have revealed that the $\nu =1$ LL edge stripe is highly incompressible while the $\nu =2$ stripe is more compressible under the conditions which seem relevant to the present experiment. 
Even if the inner ring (spin-up edge state) is incompressible, the crossing of its SP states with the Fermi level still affects the electron tunneling to/from the outer (spin-down) ring through the single electron charging effect if the outer ring is (partially) compressible.
When a SP state of the inner spin-up ring crosses the Fermi level and its electron occupation is changed, excess charge arises in the outer spin-down ring so as to screen the charge on the inner ring.
Based on our earlier work on antidot lattices~\cite{latticeJPSJ}, we infer that the effective potential slope for the $\nu =2$ LL edge on the verge of delocalization in this field range is less steep suggesting its substantially compressible nature. 
Thus we conclude that the Coulomb blockade model~\cite{Kataokadf}, which asserts that resonant tunneling to the outer ring should occur twice per $h/e$ period, is appropriate for the present case. 
As elucidated by Kataoka \textit{et al.}~\cite{Kataokadf}, partially compressible nature of the outer spin ring should lead to pairing of resonances; i.e., if the outer ring only imperfectly screens the charge on the inner ring, the exact $h/2e$ periodicity should be broken.
This is also in accordance with the present observation of the paired-dip structure shown in Fig.~\ref{FIG6}(a). 

Another noteworthy feature of the data in Fig.~\ref{FIG6}(a) is that the amplitude of the smaller dip decreases with increasing $\theta$ until it disappears at $\theta = 64^\circ$.
Rotating the sample, the Zeeman energy becomes large at the same perpendicular field. 
The energy level spacing between antidot states with opposite spins changed by the Zeeman energy causes the slight change of the separation between adjacent dips. 
The larger Zeeman energy, furthermore, leads to the weaker coupling between the outer and inner spin rings around the antidot because the separation between the LLs at the Fermi energy becomes large with increasing the Zeeman splitting. 
At $\theta =64^\circ $, the coupling is so weak that the inner spin ring cannot interact with the outer spin ring and the AB oscillation with the single $h/e$ period emerges. 


\begin{figure}[tb]
\begin{center}
\includegraphics[width=0.95\linewidth]{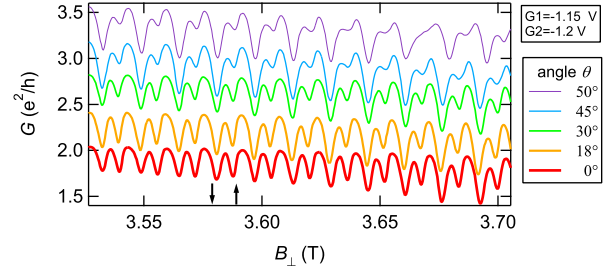}
\caption{(Color online) The conductance $G$ across the antidot as a function of perpendicular field component $B_\perp$ on the higher field side of 
$\nu_\mathrm{c}=2$ for different tilt angles $\theta $. The successive traces are offset by $0.4~e^2/h$. 
Up and down arrows indicate the dips caused by the resonances of the edge states with up and down spin, respectively. 
The source-drain bias $V_\mathrm {SD}=70~\mathrm {\mu V}$ at $\theta =0^\circ ,18^\circ , 30^\circ , 45^\circ $ and 
$50~\mathrm {\mu V}$ at $\theta =50^\circ $ are applied to cancel the offset voltages.}
\label{dfZe}
\end{center}
\end{figure}

The two sets of energy spacings plotted in Fig.~\ref {dfE}(b) (obtained from the diamonds in Fig.~\ref {dfE}(a)) correspond to $E_\mathrm {ch} + \Delta E^{(1)}_\mathrm{SP}$ (open red circles) and $E_\mathrm {ch} + \Delta E^{(2)}_\mathrm{SP}$ (solid blue circles). 
The single-particle level spacing of the outer ring $\Delta E^{(2)}_\mathrm{SP}$ is much smaller than that of the inner ring $\Delta E^{(1)}_\mathrm {SP}$ because the effective potential slope for the former is much less steep than that for the latter, as sketched in Fig.~\ref{dfSD}(c).
Assuming that $\Delta E^{(1)}_\mathrm {SP}$ and $E_\mathrm{ch}$ can be obtained by extrapolating the magnetic-field dependence shown in Fig.~\ref{FIG5}, we acquire $\Delta E^{(1)}_\mathrm {SP} \simeq 150~\mathrm {\mu eV}$ and $E_\mathrm{ch} \simeq 150~\mathrm{\mu eV}$ at $B \sim 4.2~\mathrm T$. 
These values are roughly consistent with the energy spacing $\Delta E=280~\mathrm{\mu eV}$ at $B \sim 4.2~\mathrm T$. 
The smaller diamonds (solid blue circles) disappear and merge into the larger ones at $B \sim 4.7~\mathrm T$, where the conductance $G$ begins to deviate from the $\nu_\mathrm{c}=2$ plateau value and decreases toward the $\nu_\mathrm{c}=1$ plateau value. This observation supports the presumption that the smaller diamonds represent the energy gaps of the outer ($\nu =2, \downarrow$) ring. 

Figure~\ref {dfZe} shows the conductance $G$ on the high-$B$ side of the $\nu_\mathrm{c}=2$ plateau for several values of $\theta $, taken in another experimental run after different cool-down from room temperature.
Here, the $\nu _\mathrm c=2$ state is brought to a lower $B_\perp $ range than the case in Fig.~\ref {FIG6} by applying a more negative  voltages to the side gates. 
With increasing $\theta $, one of the paired dips grows while the other diminishes. 
This behavior can be also explained by the change of the distances between the edge states with the Zeeman splitting. 
As the Zeeman splitting increases, the spin-down $\downarrow $ state around the antidot and the spin-down $\downarrow $ extended edge state become closer to each other and the dips related to down spin exhibit the larger amplitudes owing to the stronger coupling in the (2-2) tunneling process. 
While, the dips related to the up-spin exhibit the smaller amplitude because the distance between the spin-down $\downarrow $ state and the spin-up $\uparrow $ state around the antidot becomes larger, that is, the coupling between these states becomes weaker. 


Thus, on the high-$B$ side of the $\nu_\mathrm{c}=2$ QH state, the Zeeman splitting does not play a role on the energy gap between the SP antidot states with opposite spins. However, it changes the coupling between the extended edge state of the leads and the antidot state with the same spin and between the antidot states with opposite spins.  

\section{Conclusion}
We have systematically extracted the effect of the Zeeman energy on the evolution of the $h/2e$ AB oscillations in the vicinity of $\nu_\mathrm{c}=2$ using the tilted-field experiment. We have demonstrated the marked difference of the electronic states around an antidot between the two sides of the $\nu_\mathrm{c}=2$ QH state. 
On the lower field side of the $\nu _\mathrm{c}=2$ plateau, the evolution of the double-peak structure and the concomitant changes in the conduction spectra 
as a function of $V_\mathrm{SD}$ and $B_\perp$ are interpreted within the simple picture of spin-resolved single-particle states with the Zeeman splitting. 
On the higher field side of the $\nu_\mathrm{c}=2$, by contrast, the non-interacting picture of the single-particle states is no longer valid. 
In this regime, the partially compressible region of the $\nu =2$ LL around the antidot, which is caused by the delocalization and the screening effects, has a crucial role on 
the AB conductance. We argue that the principal role of the Zeeman energy in this regime is to change the coupling between the edge states with opposite spins around an antidot.

\section*{Acknowledgment}
We would like to thank Dr. M. Kataoka for helpful discussions. 
This work is supported by the Grant-in-Aid for Scientific Research (A) (18204029) from the Ministry of Education, Culture, Sports, Science and Technology (MEXT), Japan, and by Special Coordination Funds for Promoting Science and Technology.
One of the authors (M.K.) gratefully acknowledges the support by Research Fellowship for Young Scientists from the Japan Society for the Promotion of Science (JSPS).

\end{document}